\documentstyle[12pt]{article}
\def\adot{\dot \alpha}

\newfont{\bbbold}{msbm10 scaled \magstep1}
\def\com{\mbox{\bbbold C}}

\def\mink{\mbox{\bbbold M}}
\def\light{\mbox{\bbbold L}}
\def\flag{\mbox{\bbbold F}}
\def\gras{\mbox{\bbbold G}}
\def\proj{\mbox{\bbbold P}}
\def\real{\mbox{\bbbold R}}

\def\twist{\mbox{\bbbold T}}
\def\integer{\mbox{\bbbold Z}}
\def\cp{\mbox{\bbbold C}\mbox{\bbbold P}}

\newfont{\goth}{eufm10 scaled \magstep1}

\def\xz{\times}
\def\a{\alpha}
\def\b{\beta}
\def\C{\Gamma}

\def\varf{\varphi}

\def\p{\pi}

\def\th{\theta}

\def\beq{\begin{equation}}\def\eeq{\end{equation}}
\def\beqa{\begin{eqnarray}}\def\eeqa{\end{eqnarray}}
\def\barr{\begin{array}}\def\earr{\end{array}}

\begin{document}

\begin{titlepage}
\begin{flushright}
ENSLAPP-A-564/94\\
hepth@xxx/9512066\\
\today
\end{flushright}

\bigskip\bigskip\begin{center} {\bf
\Large{Twistors and Supersymmetry}}
\end{center} \vskip 1.0truecm

\centerline{\bf P.S. Howe\footnote[1]{Permanent address: Dept. of Mathematics,
King's College,
London}}
\vskip5mm
\centerline{L.A.P.P.}
\centerline{B.P. 110, 74941 Annecy-le-Vieux, France}
\vskip5mm

\bigskip \nopagebreak \begin{abstract}
\noindent
An overview is given of the application of twistor geometric ideas to
supersymmetry with particular emphasis on the construction of superspaces
associated with four-dimensional spacetime.
\vskip 2cm
\noindent
Talk given at the Leuven Conference on High Energy Physics, July 1995.
\end{abstract}

\end{titlepage}

\section{Introduction}

\noindent
Twistor geometry, suitably interpreted, is particularly well-suited
to the description of supersymmetric theories in the superspace
formalism.  There have been applications: to field
theories, notably supersymmetric Yang-Mills (SYM) and supergravity
(SG), leading to off-shell representations and to the
interpretation of on-shell constraints as integrability conditions,
and to super-particles, strings and extended objects in the context of doubly
supersymmetric formulations, that is, formulations with both
spacetime (target space) and worldsurface supersymmetry.
\par
The term ``twistor geometry'' as I shall interpret it in the
supersymmetric context embraces the three main geometrical
approaches to supersymmetry that have been studied:  chiral
supergeometry, light-like integrability or super twistor theory, and
harmonic superspace, each of these three formalisms being
particular examples of twistor supergeometries as I shall show.  In
general one can say that the twistor approach is a very valuable tool for
constructing and understanding superspace geometries and, in particular,
it clarifies the geometrical structure of harmonic
superspace.
Since (almost) all supersymmetric theories are amenable to one or more of
the above descriptions it follows that twistor supergeometry is
universal:  it underlies  (almost) all supersymmetric theories of interest.
\par
In this talk I shall focus on the construction of various
(flat-space) supergeometries associated with four-dimensional spacetime.
In four dimensions twistors are naturally associated with
conformal symmetry, a symmetry which, as is well-known (see for example
\cite{sohnius-1}), is particularly relevant in the
supersymmetric case for three reasons: firstly, massless, non-gravitational
theories of interest are classically superconformal, secondly, it is known that
there are four-dimensional superconformal quantum field theories, and thirdly,
the construction and study of supergravity theories is made much easier if one
uses the superconformal perspective.  For the most part I shall
work in complex spacetime;  this has some advantages from a formal
point of view and has some relevance in that quantum field theories
can be studied in regions of products of several copies of complex spacetime.
Moreover, there is
no loss of generality since one can easily impose reality in the
formalism.  In the case of standard twistor theory, there is still
a complex twistor space associated with real (Euclidean) space;  in the
case of harmonic supergeometry one finds that the complex geometry
of twistor space has to be replaced by $C\! R$
supergeometry, which will be briefly explained in section 4.
Finally, I shall give a list of some applications to supersymmetric
field theories.

\section{Twistors}
\noindent
Consider the equation
\begin{equation}
u^\a=-i x^{\a\adot} v_{\adot}
\end{equation}
\par
where $u$ and $v$ are two-component commuting spinors and  $x$ is a
four-vector in $2\times 2$ spinor form.  The pair $z=(u,v)$ is an
element of twistor space $\twist\cong\com^4$, and $x$ labels a point
in
complex Minkowski space $\mink \cong \com^4$.  Equation (1)
establishes a correspondence between $\twist$ and $\mink$ in the
following
way \cite{penrose}: firstly, if $z=(u,v)$ is held fixed, (1) defines a
$2$-plane
(called a
$\beta$-plane)
in $\mink$ which is totally null (all tangent vectors are
null) and
anti-self-dual (the bivector constructed from any two independent tangents is
anti-self-dual);  secondly if $x$ is
held fixed, then one can solve (1) for $u$ as a function of $v$.
Since (1) is clearly invariant under common rescalings of $u$ and $v$ this
second point of view determines a projective line $(\cp ^1)$
in projective twistor space, $\proj$, also known as a twistor line.
Thus we have the correspondence
\begin{itemize}
\item points $x\in M\ \rightarrow$ twistor lines $\underline{x}\in \proj$
\item points $[z] \in \proj \rightarrow$ $\b$-planes $\underline{z}\in \mink$
\end{itemize}
\par
We can present this a little differently as follows:
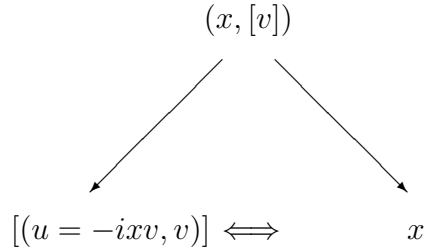
\[
\begin{picture}(300,120)(-80,-10)
\put(53,80){$(x,[v])$}
\put(60,68){\vector(-1,-1){50}}
\put(80,68){\vector(1,-1){50}}
\put(-20,0){$[(u=-ixv,v)]$}
\put(60,0){$\Longleftrightarrow$}
\put(130,0){$x$}
\put(11,50){$$}
\put(111,50){$$}
\end{picture}
\]
\par
That is, given a point $x\in \mink$ and a projective
dotted spinor
$[v]$ we can project either onto Minkowski space or onto projective
twistor space as indicated.  We can rewrite this in terms of spaces
as:
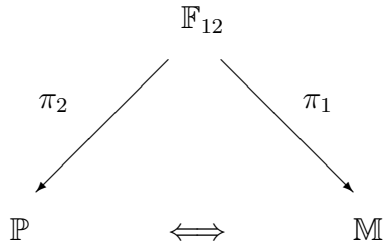
\[
\begin{picture}(300,120)(-80,-10)
\put(65,80){$\flag_{12}$}
\put(60,68){\vector(-1,-1){50}}
\put(80,68){\vector(1,-1){50}}
\put(0,0){$\proj$}
\put(60,0){$\Longleftrightarrow$}
\put(130,0){$\mink$}
\put(11,50){$\pi_2$}
\put(111,50){$\pi_1$}
\end{picture}
\]
where ${\flag}_{12}$, the space of points of $\mink$ and
projective dotted spinors, is called the correspondence space and
the projections $\pi_1$ and $\pi_2$ are given in coordinates in
the previous diagram.  This type of diagram is called a double
fibration \cite{wardwells}, and the projections should be such that $\pi_2$ is
one-to-one
on each fibre $\pi^{-1}_1(x)$ of $\pi_1$ and vice-versa, so
that the subset $\underline {x}=\pi_2 \circ\pi_1^{-1}(x)$ in $\proj$ is a
copy of the fibre $\pi_1^{-1}(x)$ and similarly for $\underline
{z}=\pi_1\circ\pi_2^{-1}(z)$, which is a subset of $\mink$.  Thus
the double fibration builds in the correspondence automatically.
We shall interpret a set of three spaces related to each other by
such a double fibration as a twistor geometry.
\par
The significance of such a geometrical set-up lies in the fact that
information about field theories on spacetime can be related to
certain (holomorphic) data on twistor space.  Most  importantly, one has the
Ward  construction for gauge theories \cite{ward,wardwells}.  Given a gauge
theory on $\mink$ we
can lift it to $\flag_{12}$ to get a gauge theory on the bigger
space which depends  trivially on the fibre coordinate, $[v]$.
Similarly, a gauge theory on ${\proj}$ will lift to a theory on
${\flag}_{12}$ which is trivial on the fibres of $\pi_2$.  Thus,
in order for gauge theories on $\mink$ and $\proj$ to be
equivalent, the former must be trivial on each fibre of $\pi_2$,
i.e. on each $\beta$-plane, while the latter must be trivial on
each fibre of $\pi_1$, i.e. on each twistor line.  More precisely,
one finds that a gauge theory on $\mink$ with vanishing
curvature on each $\beta$-plane is equivalent to a holomorphic vector bundle
over ${\proj}$ which is trivial on each twistor line.
Moreover, this result generalises to any double fibration of the
above type provided certain technical requirements are met \cite{basteast}.
\par
In gravity, the situation is slightly different. Given a (complex) spacetime
$\mink$ one wishes to know if there is a twistor space such that one can
construct a double fibration. The existence of such a twistor space
leads to constraints on the geometry of $\mink$; in standard
twistor theory these constraints involve self-duality whereas in
supergravity one can find generalised twistor spaces whose existence implies
the desired equations of constraint on the geometry of superspace.
\par
There is a systematic way of constructing double fibrations   using group
theory \cite{basteast}.  Regarding
twistor space $\twist$ as a representation space for the complex
conformal group, $SL(4,\com)$ we can construct from it a number of spaces
(7 in all) which have the following properties:
\begin{itemize}
\item They are spaces of flags in $\twist$.
\item They are homogeneous spaces of the form $P\backslash G$ where $P$ is
a parabolic subgroup of $G:=SL (4,\com)$
\item They group together in sets of 3, since $P_1\cap
P_2:=P_{12}$ is parabolic whenever $P_1$ and $P_2$ are,  so that one
automatically gets double fibrations, $P_2\backslash G\leftarrow
P_{12}\backslash G\rightarrow P_1\backslash G$, with the desired
properties.
\end{itemize}
\par
Consider the following subgroups of $G$ consisting of matrices of
the type indicated (the crosses denote non-zero elements):
\[
P_1=\left( \begin{array}{llll} \xz &\xz & & \\
                             \xz &\xz & & \\
                             \xz &\xz &\xz &\xz \\
                             \xz &\xz &\xz &\xz
\end{array}\right)
\qquad\  P_2=\left(\begin{array}{llll} \xz & & & \\
                             \xz &\xz &\xz &\xz \\
                             \xz &\xz &\xz &\xz \\
                             \xz &\xz &\xz &\xz
\end{array}\right)
\]

If we view these matrices as acting from the left on a column
vector of basis vectors of $\twist$ we see that $P_1$ leaves the plane
determined by the first two vectors invariant, whereas $P_2$ leaves
the line delivered by the first vector invariant.  Hence
$P_1\backslash G=\gras r_2(4)$, the Grassmanian of 2-planes in
$\com^4$ while $P_2\backslash G=\cp^3$.  Clearly:
\[
P_{12}=\left( \begin{array}{llll} \xz & & & \\
                             \xz &\xz & & \\
                             \xz &\xz &\xz &\xz \\
                             \xz &\xz &\xz &\xz
\end{array}\right)
\]
and the space $P_{12}\backslash G$ is the space of flags of type
(1,2) in $\com^4$, where a flag of type (1,2) is a line
embedded
in a two-dimensional subspace.
\par
All the above spaces are in fact  compact complex manifolds.  In
particular we can identify $\gras r_2 (4)$ as complexified
compactified Minskowski space $\widetilde{\mink}$.  In practice we
are interested in non-compact spacetime, and we can identify this as an
open set homeomorphic to $\com^4$ in $\widetilde{\mink}$.  This
open set can be related to a section $s(x)$ of $G\rightarrow
P\backslash G$ defined by
\begin{equation}
x\mapsto s(x)=\left( \begin{array}{lr} 1 & -ix \\
                             0 & 1
\end{array}\right)
\end{equation}
where each entry is a $2\times 2$ matrix.  By using standard
homogeneous space techniques one can easily verify  that $SL (4,\com)$
gives rise to the usual conformal group transformations of
Minskowski space.
\par
The non-compact double fibration is obtained from the compact one,
$(P_2\backslash G\leftarrow P_{12}\backslash G\rightarrow
P_1\backslash G)$, by replacing $P_1\backslash G$ by $\mink$ and
tracing this around the diagram using the projections:
\[
\begin{picture}(300,120)(-80,-10)
\put(20,80){$\p_1^{-1}(\mink):=\flag_{12}=\mink\xz\cp^1$}
\put(60,68){\vector(-1,-1){50}}
\put(80,68){\vector(1,-1){50}}
\put(-100,0){$\p_2\circ\p_1^{-1}(\mink):=\proj=\cp^3\backslash\cp^1$}
\put(60,0){$\Longleftrightarrow$}
\put(130,0){$\mink$}
\put(11,50){$\pi_2$}
\put(111,50){$\pi_1$}
\end{picture}
\]
\par

\section{Supersymmetry}
\noindent
\par
It is straightforward to generalise the foregoing to the
supersymmetric case.  $N$-extended super twistor space $\twist_{N}$ is
$\com^{4|N}$, the complex super vector space with four even
and $N$ odd dimensions;  it is   a representation space of the
complex superconformal group $SL (4|N; \com)$.  Starting from
$\twist_N$ we can construct a large number of flag supermanifolds \cite{manin}
by
looking at the parabolic subsupergroups of $G=SL(4|N;\com)$.
\par
One of the simplest examples is $ N=1$ (compactified) super Minkowski space,
${\widetilde{\mink}}$ (i.e. the body is compact), which corresponds to the
subgroup of $SL(4|1;\com)$
consisting of matrices of the form
\[
 \left( \begin{array}{cccc|c}
        \times & \times & & &  \\
        \times & \times & & &  \\
\times & \times & \times & \times & \times  \\
\times & \times & \times & \times & \times  \\
\hline  \times & \times & & & \times
\end{array} \right)
\]
\par
We identify non-compact super Minkowski space with an open set in
$\widetilde{\mink}$ in a similar fashion to the bosonic case.  In
local
coordinates $z=(x,\th,\varf)$ we have
\begin{equation}
z\mapsto s(z)=\left( \begin{array}{cr|r}
1 & -i X & -i\th \\
0 & 1 & 0\\
\hline
0 & -i\varf & 1
\end{array}\right)
\end{equation}
where $X^{\alpha {\dot \alpha}}=
x^{\a\adot}-{i\over2}\theta^\alpha\varphi^{\dot
\alpha}$.  Left and right chiral superspaces, $\mink_L\  ({\mink_R})$ are
associated with the subgroups of matrices of the forms (for $N=1$):
\[
 \left( \begin{array}{cccc|c}
        \times & \times & & &  \\
        \times & \times & & &  \\
\times & \times & \times & \times & \times  \\
\times & \times & \times & \times & \times  \\
\hline  \times & \times & \xz&\xz & \times
\end{array} \right) \qquad\hbox{\rm and}\qquad
\left( \begin{array}{cccc|c}
        \times & \times & & &\xz  \\
        \times & \times & & & \xz \\
\times & \times & \times & \times & \times  \\
\times & \times & \times & \times & \times  \\
\hline  \times & \times & & & \times
\end{array} \right)
\]
\par
so that we have the double fibration
$\mink_L\leftarrow \mink\rightarrow\mink_R$ \cite{manin}.
This is in fact an exceptional case in the sense that the space of
interest, $\mink$, is at the ``top'' of the diagram rather than at
the bottom right. The above can be extended in an obvious way to $N$-extended
super Minkowski space and the corresponding chiral superspaces. We shall denote
$N$-extended super Minkowski space by $\mink_N$, or simply by $\mink$.

Further examples of superspaces which have been used in field
theory fall into two classes:
\begin{itemize}
\item Super twistor spaces \cite{ferber}:  these have bodies which are
ordinary twistor spaces, i.e. one of the 7 spaces of the previous
section.
\item harmonic superspaces \cite{gikos, klr}:  these have bodies of the form
Minkowski space $\times$ internal flag space.
\end{itemize}
\par
The main example of supertwistor theory is given by the
double fibration \cite{witten1,harnad} $\light\leftarrow\gras\rightarrow\mink$
where the subgroups  defining ${\gras}$ and ${\light}$ consist of matrices of
the forms
\[
 \left( \begin{array}{cccc|ccc}
        \times & & & & & &  \\
        \times & \times & & & & & \\
\times & \times & \times &  & \times &\xz &\xz \\
\times & \times & \times & \times & \times &\xz &\xz \\
\hline  \times & \times & &  &\xz &\xz &\xz \\
\times & \times & &  &\xz &\xz &\xz \\
\times & \times & &  &\xz &\xz &\xz
\end{array} \right) \qquad\hbox{\rm and}\qquad
\left( \begin{array}{cccc|ccc}
        \times & & & & & &  \\
        \times & \times &\xz & & \xz&\xz &\xz \\
        \times & \times & \times &  & \times &\xz &\xz \\
        \times & \times & \times & \times & \times &\xz &\xz \\
\hline  \times & \times & \xz&  &\xz &\xz &\xz \\
        \times & \times & \xz&  &\xz &\xz &\xz \\
        \times & \times & \xz&  &\xz &\xz &\xz
\end{array} \right)
\]
\par
respectively (illustrated for $N=3$).  The space ${\light}$,
sometimes called super ambitwistor space, is the space of complex
super light-like lines in $\mink$. In the real case this is
replaced by
\[
\begin{picture}(300,120)(-80,-10)
\put(55,80){$M\xz \cp^1$}
\put(60,68){\vector(-1,-1){50}}
\put(80,68){\vector(1,-1){50}}
\put(0,0){$L$}
\put(60,0){$\Longleftrightarrow$}
\put(130,0){$M$}
\put(11,50){$\pi_2$}
\put(111,50){$\pi_1$}
\end{picture}
\]
The space $M\times \cp^1$ is called light-cone harmonic
superspace \cite{sokat}, while $L$, the space of real super light-like lines
in super
Minkowski space,
$M$, is a real subspace of dimension $(5|2N)$ of projective
supertwistor space.
\par
Harmonic superspaces \cite{gikos,hh1} are classified by a pair of integers
$p,q$
where $p +q\leq N$ and we suppose that $p,q\geq 1$.
\par
The appropriate subgroups for this case consist of matrices of the form
\[
\begin{picture}(300,200)(20,-100)
\put(0,0){$
 \left( \begin{array}{cccc|ccccccc}
        \times & \times & & & \bullet&\bullet &\bullet & & & & \\
        \times & \times & & & \bullet& \bullet&\bullet & & & & \\
\times & \times & \times & \times & \times & . & . & . & . & . & \times \\
\times & \times & \times & \times & \times & . & . & . & . & . & \times \\
\hline  \times & \times & & & \times & . & \times & & & &\\
        .      & .      & & & .      & . & .      & & & &\\
        .      & .      & & & \times & . & \times & & & &\\
        .      & .      & & & \times & . & .      & . & \times & &\\
        .      & .      & & & .      & . & .      & . & .      & &\\
        .      & .      & & & \times & . & .      & . & \times & &\\
       \xz     & \xz     &\bullet &\bullet & \times & . & .  & . & . & . &
\times \\
       \xz     & \xz      &\bullet &\bullet &.& . & .  & . & . & . & . \\
\times & \times & \bullet& \bullet& \times & . & .  & . & . & . & \times
\end{array} \right)  $}
\put(200,15)
{$ \left. \phantom{\begin{array}{c} \times \\ .
\\ \times \end{array} } \right\}p $}
\put(200,-72)
{$ \left. \phantom{\begin{array}{c} \times \\ .
\\ \times \end{array} } \right\}q $}
\end{picture}
\]
The space determined by the subgroup with no bullets is (compactified)
$(N,p,q)$
harmonic superspace, $\widetilde{\mink}_N(p,q)$, whereas the space
determined by the subgroup including the  bullets is called (compactified)
analytic
$(p,q)$ superspace, $\widetilde{\mink}_{NA}(p,q)$.  These two spaces
form a double fibration with super Minkowski space:
$\widetilde{\mink}_{NA}(p,q)\leftarrow\widetilde{\mink}_N(p,q)\rightarrow
\widetilde{\mink}_{N}$.
In the non-compact case $\mink_N(p,q)=\mink_N\times {\flag}_{p,N-q}(N)$ where
the internal flag space (${\flag}$ for short in the following)
is represented by the bottom right part of the above diagram.

The correspondence is between points of $\mink$ and copies of
${\flag}$ in $\mink_{NA}$ and between points of analytic space
and $\triangle (p,q)$ planes in $\mink$.  The latter are planes
of dimension $(0|2 (p+q))$ which have $2p$ tangent vectors with  undotted
indices
(of the form $X^{\alpha i}, i=1,...N$) and $2q$ tangent vectors
with dotted indices (of the form $Y^{\dot\alpha}_i$) such that
$X^{\alpha i}Y^{\dot\alpha}_i=0$ for any such pair of vectors.  A
basis for the tangent spaces to any such plane is given by the derivatives
\beq
D_{\a r}=u_r{}^i D_{\a i} \qquad D_{\adot}^{r'}=v_i{}^{r'} D_{\adot}^i
\eeq
where $r=1,\dots p;\ r'=(N-q+1)\dots N$, the $D's$ are the usual superspace
covariant derivatives and where the matrices $u,v$ have maximal rank and
satisfy
\beq
u_r{}^i v_i{}^{r'} =0.
\eeq
It is not difficult to see that this equation defines the flag
manifold ${\flag}_{p,N-q}(N)$.
A field, $A$, on $\mink_N(p,q)$ is called $G$-analytic if it
satisfies the generalised chirality constraints
\beq
D_{\a r} A= D_{\adot}^{r'} A =0.
\eeq
Such a field is equivalent to a field on analytic superspace.

\section{Reality}
\noindent
\par
The prototype twistor double fibration collapses to a single
fibration of the form  $\real^4\times\cp^1\rightarrow
{\real}^4$ when Minkowski space is replaced by Euclidean space, $\real^4$; this
is the
starting point relevant to the  study of self-dual Yang-Mills theory in
Euclidean space.  In a similar manner, it is often convenient to
consider only single fibrations in the case of the chiral and harmonic
supergeometries appropriate for real super Minkowski space.
However, whereas in the case of Euclidean twistors the twistor
space $Z=\real^4\times\cp^1$ is naturally a complex space
(it is ``almost'' $\cp^3$), in the supersymmetric case it is
necessary to generalise the notion of a complex structure to that
of a $C\! R$ structure \cite{rs}.
\par
Let $M$ be a real $(2n +m)$-dimensional (super)manifold where
$n,m\in {\integer}\  ({\integer}^2$ in the supercase),  and let $T_c$
be the complexified tangent bundle.  A $C\! R$ structure (of rank $n$)
on $M$ is an $n$-dimensional sub-bundle of $K$ of $T_c$ such that
\begin{itemize}
\item $K_p\cap\bar K_p=0, \ \forall p\in M$
\item $X,Y\in \C(K)\Rightarrow [X,Y]\in \C(K)$
\end{itemize}
where $\bar K$ denotes the complex conjugate of $K$, $K_p(\bar K_p)$ the
fibre of $K(\bar K)$ at $p\in M$, and $\Gamma (.)$ denotes the space of
sections
of a bundle.  On a $C\! R$ (super)manifold
there is a generalisation of the Dolbeault operator $\bar\partial$ on a complex
manifold which is
denoted by $\bar\partial_K$ and defined  on scalars by $\bar\partial_K
f=\pi \circ df$ where $\pi$ denotes the projection: one-forms
$\rightarrow$ sections of $\bar K^*$, the dual space of $\bar K$.  A
$C\! R$ analytic function is one which satisfies $\bar\partial_K f=0$.
A simple example of the above is given by real $N=1$ superspace $M$
which is a $C\! R$  supermanifold of rank $(0|2)$ with $\bar\partial_K$ being
simply
the operator $\bar D_{\dot\alpha}$.
\par
Real harmonic superspace $M_N(p,q)$ is $M_N\times {\flag}$ where ${\flag}$ is
the same internal complex flag manifold as in the previous
section but which can also be thought of as the coset space $S((U(p)\times
U(N-p-q)\times U(q))\backslash SU(N)$.  (Thus a field on $M_N(p,q)$ can be
expanded in $SU(N)$ harmonics on $\flag$ with coefficients which are ordinary
superfields, whence the nomenclature). It
is a $C\! R$ supermanifold of rank $(\dim_c \flag\vert 2(p+q))$.  The
components of the $C\! R$ operator $\bar\partial_K$ are $\{\bar\partial
_F,D_{\alpha r}, D^{r'}_{\dot\alpha}\}$ where
$\bar\partial_F$ is the usual Dolbeault operator on $\flag$.  In
addition one can have $G$-analytic fields which are annihilated by
the odd derivatives $D_{\alpha r}, D^{r'}_{\dot\alpha}$, but which
need not be analytic on ${\flag}$.
\par
\section{Applications}
\noindent
\par
1) Massless on-shell supermultiplets can in many cases be described
by $C\! R$ analytic superfields in appropriate harmonic superspaces \cite{hh1}.
For example, $N=8$ linearised SG is described by a single component
superfield, $W$, on $M_8(4,4)$ superspace.  The $N=8$ linearised
three-loop
counterterm \cite{kallosh,hst} can be written in the manifestly supersymmetric
form $\int d\mu W^4$ where $d\mu$ is
the measure for $G$-analytic fields on $M_8(4,4)$, (generalised
chiral measure) \cite{hh1}.
\par
2) SYM theories.  For $N=2,3$ the superspace constraints of SYM
\cite{sohnius} may be
described by applying the Ward construction to either the super
twistor \cite{witten1} or harmonic \cite{gikos,gikos2} double fibrations
(complex
case);  that is, we
have the double double fibration:
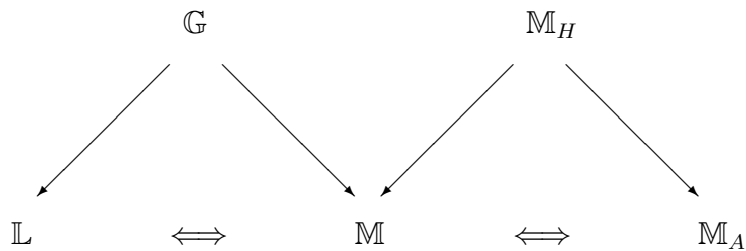
\[
\begin{picture}(300,120)(-80,-10)
\put(5,80){$\gras$}
\put(135,80){$\mink_H$}
\put(0,68){\vector(-1,-1){50}}
\put(20,68){\vector(1,-1){50}}
\put(130,68){\vector(-1,-1){50}}
\put(150,68){\vector(1,-1){50}}
\put(-60,0){$\light$}
\put(0,0){$\Longleftrightarrow$}
\put(130,0){$\Longleftrightarrow$}
\put(70,0){$\mink$}
\put(200,0){$\mink_A$}
\end{picture}
\]
where $\mink_H\ (\mink_A)$ denotes the relevant harmonic (analytic)
superspace.  In the real case, it appears that only the harmonic
formalism can be used to write actions \cite{gikos,gikos2}. For $N=1$ the
supertwistor diagram
is still relevant, but the chiral double fibration can also be
used.
\par
3) Superconformal geometry.  In complex superspaces, superconformal
transformations can be defined as those transformations of the
correspondence space that leave the double fibration invariant \cite{hh2}.  In
the case of real chiral and harmonic superspaces, this reduces to the
preservation of the appropriate $C\! R$ structures, although there are
some subtleties \cite{gikos3,hh2}.  Deformations of these $C\! R$ structures
(for
$N\leq4$) determine off-shell field
representations of conformal supergravity (CSG).
\par
4). Non-linear CSG.  The $C\! R$ structure of $N=1$ superspace gives a global
description of Ogievetski-Sokatchev supergeometry \cite{os,siegel}.  For $N\ge
2$,
the geometry of CSG \cite{howe1,gikos4} resembles quaternionic geometry
\cite{salamon} in the
non-supersymmetric case.  In this latter case one has a complex twistor
space, $Z$, which is a fibre bundle over the quaternionic manifold
$M$ with fibre $\cp^1$, whereas in the harmonic superspace
description of CSG, harmonic superspace is a $C\! R$ supermanifold as well as
being
a fibre bundle
which fibres over superspace $M$ with fibre the flag manifold,
${\flag}$ \cite{hh1}.
\par
5) Extended Poincar\'e SG.  The basic constraints of on-shell extended
supergravity theories are superconformal \cite{howe2} and can again be obtained
from the perspective of the $C\! R$ supergeometry of harmonic
superspace, at least for $(p,q) = (1,1)$ \cite{hh1}.
\par
6) Lower dimensional spacetimes.  For spacetime dimensions
$d=1,2,3$, constructions similar to those given here can be carries
out using the appropriate superconformal groups $OSp(N|1)$ and
$OSp(N|2)$ \cite{hl,leeming}, see also \cite{zupnik}.
\par
7) Higher dimensional spacetimes.  With the exception of $d=6$ the
superconformal approach is not so useful in dimensions greater than
4.  Instead twistors make their appearance via the interpretation
of the $d$-dimensional Lorentz group $SO(1,d-1)$ as the conformal
group of $(d-2)$-dimensional Euclidean space.  When combined with
supersymmetry this has applications via lightlike integrablity \cite{light} and
pure spinors \cite{pure}
to higher dimensional field theories as well as to supersymmetric
particles \cite{part} and extended objects \cite{obj}.
\par
8) Euclidean and $(2,2)$ signatures. It is possible to impose reality on the
formalism in other ways so as to construct superspaces associated with
Euclidean
space or with four-dimensional space with a (2,2) signature metric. In these
situations, self-duality can be supersymmetrised and supertwistor techniques
applied. See, for example references \cite{euclid}.

9) Quantum supersymmetry.  Up to now the applications of the above
formalism to quantum supersymmetry have been to perturbation theory using
off-shell
superfield formalisms (where applicable) and  superspace
Feynman diagrams.  However, there are indications that there may be
direct applications to non-perturbative quantum field theory.  Work
on this is currently in progress \cite{hw}.

\end{document}